\documentclass[showpacs,aps]{revtex4}

\usepackage{graphicx}  

\def\bea{\begin{eqnarray}}
\def\eea{\end{eqnarray}}
\def\ben{\begin{equation}}
\def\een{\end{equation}}
\def\benu{\begin{enumerate}}
\def\enu{\end{enumerate}}

\def\n{n}

\def\sss{\scriptscriptstyle\rm}

\newcommand{\h}{\hat{H}}

\def\half{\frac{1}{2}}

\def\c{_{\sss C}}
\def\C{_{\sss C}}
\def\s{_{\sss S}}

\def\H{_{\sss H}}
\def\xc{_{\sss XC}}

\def\hx{_{\sss HX}}
\def\HX{_{\sss HX}}
\def\hxc{_{\sss HXC}}

\def\ext{_{\rm ext}}

\def\up{_\uparrow}
\def\dn{_\downarrow}

\def\PR{Physical Review\ }
\def\PRA{Physical Review A\ }
\def\PRB{Physical Review B\ }
\def\PRL{Physical Review Letters\ }
\def\JCP{J. Chem. Phys.\ }

\newcommand{\csch}{\;\mbox{csch}\;}
\def\dilog{\Re\mbox{DiLog}}

\def\1dh2{1D H$_2$}

\begin{document} 
\headheight 50pt


\title{Ground and excited-state fermions in a 1D double-well, exact and time-dependent density-functional solutions }
\author{R. J. Magyar}  
\affiliation{The National Institute of Standards and Technology, 100 Bureau Dr. MS 8380, Gaithersburg, MD 20899}


\begin{abstract} 

Two of the most popular quantum mechanical models of interacting fermions are compared 
to each other and to potentially exact solutions for a pair of contact-interacting 
fermions trapped in a 1D double-well potential, a model of atoms in a quasi-1D optical 
lattice or electrons of a Hydrogen molecule in a strong magnetic field.  An exact 
few-body Hamiltonian is solved numerically in momentum space yielding a 
highly-correlated eigenspectrum.  Additionally, approximate ground-state energies are 
obtained using both density functional theory (DFT) functional and 2-site Hubbard 
models.  A 1D adiabatic LDA kernel is constructed for use in time-dependent density 
functional theory (TDDFT), and the resulting excited-state spectrum is compared to the 
exact and Hubbard results.  DFT is shown to give accurate results for wells with small 
separations but fails to describe localization of opposite spin fermions to different 
sites. A \emph{locally cognizant} (LC) density functional based on an effective local 
fermion number would provide a solution to this problem, and an approximate treatment 
presented here compares favorably with the exact and Hubbard results.  The TDDFT 
excited-state spectrum is accurate in the small parameter regime with non-adiabatic effects 
accounting for any deviations.  As expected, the ground-state Hubbard model outperforms DFT at 
large separations but breaks down at intermediate separations due to improper scaling 
to the united-atom limit.  At strong coupling, both Hubbard and TDDFT methods fail to capture the appropriate energetics.

\end{abstract}

\date{\today} 
\pacs{
31.15.Ew,  
71.15.Mb,  
71.10.-w,  
}  

\maketitle


\section{Introduction}
\label{s:intro} 

One of the underlying challenges in computational physics, notably in density 
functional theory, is the accurate and reliable treatment of many-body interactions 
in matter.  The typical interaction-type that most electrons in solids experience is 
the Coulomb interaction, and this is a pernicious one owing to its long-ranged nature.  
Many of the difficulties in providing accurate numeric results for Coulomb-interacting 
systems are the results of the long-range.  It would be insightful and beneficial to the 
development of techniques that explicitly describe many-body systems if the Coulomb 
interaction were instead a more local version.  One such replacement, the contact 
interaction offers a powerful test case to probe specifically the many-body problem 
without long-range complications.  This is particularly important in the analysis and 
formal improvement of practical methods of electronic structure theory such as Hubbard 
theories and density functional ones.  In this paper, we explore a simple model system 
of fermions that interact via a contact interaction.  Remarkably, the interacting 
quantum problem is shown here to reduce to a solvable system of integral equations.  
Additionally, we extend developments in the density functional modeling of 
contact-interacting systems to the time-dependent domain.  This work has implications 
both for the formal development of density-functional theory and for the practical 
numeric analysis of quasi 1D systems.

The contact interaction can arise as a simplified form of 3D interactions in highly 
confined quasi-1D systems, experimentally realizable in 1D optical lattices 
\cite{CML6,SMG6,MSG5}.  In these situations, the contact term is a reasonable 
approximation to the interaction when the ratio of the interaction strength to the 
transverse well width is large.  An additional approximation made is that the 
wave-function, although still 3D, factors into transverse and longitudinal components 
with the interaction only affecting the longitudinal part. This decomposition is valid 
when the transverse and longitudinal length scales are drastically different.  We 
consider only the nontrivial longitudinal part.  Thus, the 3D interaction is replaced 
by a 1D one. Admittedly, a more realistic reduction of the 3D Coulomb problem to an 
effectively 1D one is widely debated, and, here, we have chosen a particular form 
partially motivated by the advantages it offers computationally.

One of the computational advantages of the $\delta$-function interaction within the 
DFT formalism is that the exchange and Hartree density functionals are explicit 
functionals of the fermion density, a result implying that the local-density 
approximation (LDA) to exchange is self-interaction free.  Thus, the functionals 
used for Hartree and exchange represent the exact-exchange formalism (EXX).  This 
locality of EXX is not true for the long-ranged 3D Coulomb interaction, and performing 
EXX in 3D is a significantly more complicated endeavor.  For the full LDA in the 1D case, 
any inaccuracies are caused by the mismodeling of correlation. In 3D, the 
inaccuracies are mixed between exchange and correlation due to the long-ranged nature of 
the Coulomb potential.  A notable example of a long-ranged correlation problem is the 
inability of the LDA to localize single electrons on distant sites.  Hubbard theory, 
on the other hand, preserves this limit but sacrifices accuracy.

For this reason and the clarifying simplicity of 1D models, the $\delta$-function 
model for 1D fermions has been considered in several recent studies.  In particular, 
the introduction of a local density correlation functional has provided exceptionally 
accurate results for the ground-state of finite systems such as the Diracium 
\cite{MB4} and the 1D analog of Hooke's atom \cite{MB4,XPAT6}.  A modified 
parameterization of the local density approximation (LDA) has been used to describe 
interacting fermions in harmonic confining potentials \cite{XPAT6}. A time-dependent 
Thomas-Fermi theory has been used to predict the excited-state properties of a 
many-fermion system but did not amount to the Kohn-Sham (KS) formulation of DFT 
\cite{KZ4}.  More general problems involving this contact interaction also have been studied recently \cite{ABGP4,BBD6,F4}.

In this paper, we investigate a two-site $\delta$-function problem, the 1D quantum 
analog of the H$_2$ molecule \cite{NB6} or a 2-site optical lattice, hitherto referred 
to as \1dh2.  The author introduces a 1D contact-interacting time-dependent 
density functional theory (TDDFT) and compares the results to the exact solutions and 
to the 2-site Hubbard results.  In order to accurately express the 1D adiabatic LDA 
kernel, it is found necessary to improve the existing parameterization of the result for the reference system 
to include higher-order terms at the low- and high-density limits.  The resulting 
functional reproduces the previously known expansion terms reported in Refs. 
\cite{MB4,XPAT6} but additionally includes a critically important high density term 
needed to describe the adiabatic LDA kernel.  The author develops a numeric scheme to 
obtain the exact bound-state spectrum for \1dh2 from a set of 1D integral equations.  
The results are tested against a general form of the virial theorem.  Numeric 
integration techniques are developed to handle the oscillatory nature of the 
integrand.  In the stretched system, the author contrives a scheme to correct the 
long-range correlation self-interaction error.  The results are compared to the 
well-known two-site Hubbard model that is expected to be accurate in the 
long-separation limit.

Throughout, we assume that our 1D fermions have the same mass as electrons, and we use 
atomic units ($e^2=\hbar=m_e=1$) so that all energies are in Hartrees and all lengths 
in Bohr radii.  A realistic correspondence between the interaction strength $\lambda$ 
and the details of a trapping potential is given by Ref. \cite{G97}.  Here, the 
external potential and interaction strengths are arbitrarily chosen so that the total 
energies come out in the typical chemical range for the sake of physical intuition.


\section{One-dimension contact-interacting Fermions in a double-well}
\label{s:1dh2} 

In this section, we state the quantum mechanical model of trapped 1D fermions.  From 
the non-interacting case, we anticipate how the many-body wave-functions are 
structured.

The 1D Hamiltonian for the particle-pair in a 1D well is 
\bea
\h=
-\half \sum_{i=1}^2 \frac{d^2}{dx_i^2} 
+\lambda \delta(x_1-x_2) 
- Z \sum_{i=1,\pm}^2 \delta(x_i\pm a)  . \nonumber\\
\label{1dh2}
\eea
The terms from left to right are the kinetic term, the contact interaction strength, 
and the trapping potential with $a$ being half the inter-well spacing.  $Z$ and 
$\lambda$ are the relative strength of the local and interaction potentials.  $i$ 
labels the fermion number, and $x_i$ is the position of the $i$-th fermion.  The lowest energy eigenfunction or ground-state wave-function is assumed to vanish infinitely far from the trapping potential.

The non-interacting problem has two single-particle eigenvalues: $\epsilon_\pm = - 
k_\pm^2 /2 $ with $k_\pm= Z+ 1/(2a) \mbox{LambertW} \left(\pm 2 a Z \exp(-2 a 
Z)\right)$\cite{BBD6,solvedft}.  The `$+$' corresponds to the lower energy spatially 
symmetric (\emph{gerade}) single-particle bonding state and the `$-$' corresponds to 
the anti-symmetric (\emph{ungerade}) single-particle anti-bonding state.  The 
$\mbox{LambertW}(y)$ function is the principle solution for $x$ of $y=x\exp(x)$.  The 
normalized single particle wave-functions are $\phi_\pm(x) = N_\pm \left( 
\exp(-k_\pm|x-a|)\pm\exp(-k_\pm|x+a|)\right) $ where $x$ is a non-interacting 
particle?s position, and $N_\pm$ equals $1/\sqrt{\frac{2}{k_\pm}\pm\left(4 
a+\frac{2}{k_\pm}\right)\exp(-2 k_\pm a)}$.  While a \emph{gerade} ground-state exists 
for all values of $aZ$, the \emph{ungerade} state only exists for $aZ>1/2$.  We refer 
to the large separation case as stretched $H_2$ and the small separation as crushed 
H$_2$ tending to the united-atom limit. Table \ref{t:spec} characterizes the various 
non-interacting two-particle states and defines what is meant by the term 
designations, $S_0$, $S_1$, $S_2$, and $T_1$.

The virial theorem is important in practical applications where it is often used to 
verify numeric results.  In its fundamental form, it states that $2 \langle \hat{T} 
\rangle = \langle \hat{x} \frac{d}{dx}V(\hat{x}) \rangle$.  The expectation value denotes an average over spatial variables of the respective eigen-function solutions of Eq. \ref{1dh2}.  The modified version, 
generated through integration by parts and typically used for the Coulomb interaction, 
is $2 \langle \hat{T} \rangle = -\langle V(\hat{x}) \rangle$ but does not hold true 
here due to the fixed location of the external potential.  A careful integration by 
parts shows that a generalized version,
\bea 
2\langle \hat{T} \rangle + \langle V(\hat{x}) \rangle = -2 a \int_{-\infty}^{\infty}dy\; \psi^\ast( a,y) \overline{d/da \psi(a,y)} \nonumber \\ +2 a \int_{-\infty}^{\infty}dy\; \psi^\ast(-a,y) \overline{d/da \psi(-a,y)} , \label{virial} 
\eea 
is valid for the ground-state of the system. Note that we use $y$ her for the spatial 
coordinate to avoid confusion with $x_1$ and $x_2$.  $\overline{d/da \psi(a,y)}$ is 
the average of the right and left derivatives with respect to the first argument. The interaction potential part needs no 
modification.  A many-site version of Eq. \ref{virial} holds for multiple 
$\delta$-wells but has more terms on the right of the equal sign.  The right hand side 
of the general version vanishes for an isolated $\delta$-well and a periodic lattice 
of wells because in these cases the average derivative is null by symmetry, and the form of the virial theorem, $2 \langle \hat{T} \rangle = -\langle V(\hat{x}) 
\rangle$, is justified.

In the interacting problem, there are only two relevant parameters, $\lambda$ and $a$.  
The external potential strength can be scaled to unity leaving the energy in units of 
$Z^2$ times the atomic unit and length in units of $1/Z$ times the atomic unit.  This 
leaves four regimes to consider.  $\lambda$ small and $a$ large is the 
weakly-interacting two well solution. $\lambda$ large and $a$ small is the highly 
correlated double well solution.  For sufficiently small $a$, the wells merge to the 
united-atom limit.  When $\lambda$ and $a$ are large, we have a system where 
long-range correlations can be important.  Finally, $\lambda$ small and $a$ small is a 
regime where the interaction can be treated perturbatively, and the well is almost a 
single well.  Large $a$ is the domain of validity of the Hubbard model.

This paper examines 2 of these regimes in detail: the small width double-well, $a=1$, 
and the large range hopping scale, $a=2$, both with $Z=1$.  We note that in the former 
range, the separation between the wells is large enough to exceed the united atom 
limit yet still preserves the existence of at least two bound states.  The latter case 
($aZ=2$) is the regime where Hubbard theory is designed to be maximally valid.  For 
the remainder of the paper, we assume that $Z=1$.

The two-site Hubbard will be compared to the more general density functional and exact 
solutions.  The Hubbard model itself is important for the conceptual picture it 
provides and its role in the popular LDA+U method \cite{AAL97,SK0}.  Furthermore, the 
Hubbard picture is designed to reproduce localization in the stretched limit.  A 
property that is not matched in the LDA version of DFT. In the Hubbard approach, the 
Hamiltonian Eq. \ref{1dh2} is simplified to a Hubbard Hamiltonian \cite{ZGJB97}.  The 
simplification is valid when the wave-function overlap between sites is small as is 
the case for $a=2$ but not $a=1$.  The approximate Hamiltonian can be diagonalized 
exactly and relies on two parameters, $t$ (the hopping term) and $U$ (the on-site 
repulsion).

To find $t$, the hopping term, we consider linear combinations of the non-interacting 
single particle orbitals that give left and right localized fermions: $\phi_R(x)= 
(\phi_+(x)+\phi_-(x))/\sqrt{2}$ and $\phi_L(x)= (\phi_+(x)-\phi_-(x))/\sqrt{2}$.  The 
hopping term is the projection, on one localized wave-function, of the kinetic energy 
operator acting on the other localized fermion wave-function: $t=1/2 \int dx\; 
\phi_R(x) \nabla^2 \phi_L(x) =1/4(k_+^2-k_-^2)$.  Note that we defined $t$ to be positive.  This definition is chosen to 
reproduce the proper long range limit. Traditionally, the $t$ term is fixed using 
non-orthogonal localized solutions.  The difference is negligible in the large 
separation limit.  For the small separation limit, our definition of $t$ gives a 
different result than the traditional hopping term.  This is because the $\phi_-$ 
state is no longer bound for small $a$, and consequently, the maximally 
\emph{localized} solutions are not really localized to any one site.  Curiously, the 
convention used here more accurately describes the non-interacting case at small 
separations than the traditional definition.

The Hubbard Hamiltonian can be exactly diagonalized and has eigenvalues $E_{S_0}= 2 
\epsilon + \frac{U}{2} - \half \sqrt{ 16 t^2 + U^2}$, $E_{S_1}= 2 \epsilon + U$, and 
$E_{S_2}= 2 \epsilon + \frac{U}{2} + \half \sqrt{ 16 t^2 + U^2}$.  Physically, the 
first excited singlet state at large separation, $a$, corresponds to 2 fermions 
localized on one site.  This is the Diracium system and a distant empty site.  If we 
make a correspondence between the Hubbard energy of $S_1$ and the total energy of 
Diracium, $U$ can be expressed as a function of $\lambda$.  The total energy for $S_1$ 
is written $E = 2\epsilon + U$ with $\epsilon<0$.  The 2-particle allowable range for 
$U$ from Diracium is $0$ to $-\epsilon$.  Above this limit, at $\lambda_{crit.}$, 
$U^\infty_{crit.}=-\epsilon$, and the single well no longer binds two fermions.  We 
match the Hubbard $U$ value to give exact results given in Refs. \cite{R71} and 
\cite{MB4}.  For example, a $U$ of $0.354$ gives the correct energy for Diracium with 
$\lambda=1$ and $Z=1$. For numerical convenience, we parameterize the interaction 
energy versus $\lambda$ when $Z=1$: $U(\lambda)\approx 0.500 \lambda-0.163 \lambda^2 
+0.017 \lambda^3 +{\cal O}(\lambda^4)$, valid for $\lambda<\lambda_{crit}$ with a max 
error of about 0.5 \%.

Despite scaling problems, the Hubbard model is well trusted at large $a$ because its 
computational convenience and its facility with handling long-ranged ground-state 
correlations.  LDA does not handle these correlations, and even the exact solution 
method must be carefully formulated in this limit.


\section{Time-Dependent Density Functional Theory}
\label{s:tddft}

In this section we generalize work, done by us and others, in ground-state 
density-functional theory to the time-dependent case.  This extension requires several 
improvements upon the ground-state theory.  There are four main results of this 
section: 1. the discovery of the importance in TDDFT of the third term in the high 
density correlation energy, 2. the creation of an improved correlation functional that 
properly describes this term, 3. the proposal of a local effective fermion measure 
based on KS orbitals, and 4. the introduction of a new correlation functional that can 
properly describe spatially separated systems.  The results can be found in equations 
\ref{echd}, \ref{epsclda}, \ref{localparam}, and \ref{toggle} respectively.

A general approach to find the ground-state energy and fermion density of  \1dh2  is to use Density 
Functional Theory (DFT).  
In ground-state DFT, the details of the 
external potential are kept, but the many-body interaction is transformed to an 
effective local potential derived from the exchange-correlation density functional 
\cite{HK64,KS65}.  Given the exact exchange-correlation functional, DFT would return 
the exact results for the total energy and density.  In practice, this 
exchange-correlation contribution is approximated.  An active area of research in the 
3D case is to improve the accuracy and reliability of approximations to the 
exchange-correlation functional.

According to spin-density functional theory \cite{HK64}, the ground-state total energy 
is a functional of the particle density and the local magnetization.  In this work, we 
make the assumption that an axis of magnetization is chosen, and the local 
magnetization is given by $\zeta(x) = 
(n_\uparrow(x)-n_\downarrow(x))/(n_\uparrow(x)+n_\downarrow(x))$ where $n_\uparrow$ 
and $n_\downarrow$ are up- and down-spin densities projected on the magnetization 
axis.  The total ground state energy can then be decomposed as follows: $E[\n,\zeta] = 
T\s[\n,\zeta] + E\H[\n] + E\xc[\n,\zeta] + \int d x \; v\ext(x) \n(x)$ in 1D where 
$E\H[\n]$ is the exactly known Hartree or classical density-density interaction 
contribution, $v\ext(x)$ is the given inhomogeneous potential, $T\s[\n,\zeta]$ is the 
exactly known kinetic energy of non-interacting fermions at a given density, 
$E\xc[\n,\zeta]$ is the exchange-correlation energy.  The solution for the problem of 
interest is found by studying the Kohn-Sham (KS) system, the non-interacting 
counterpart to the physical system \cite{KS65}. The spin-densities are obtained from 
the occupied KS orbitals, $n_\sigma(x) = \sum_{i,\mbox{occ.}}
|\phi_{i,\sigma}(x)|^2  $.

In the contact 1D case, the Hartree and exchange terms, $E\HX[n\up,n\dn] = \lambda 
\int dx\; n\up(x)n\dn(x)$, are known exactly, and only the correlation energy 
functional must be approximated in practice.  It is important to note that the Hartree and exchange 
functional for this contact interaction is self-interaction free, so in essence, the pure density functional already includes explicitly exact-exchange (EXX).  Many of the known 
practical limitations of 3D DFT can be addressed by applying EXX but in 3D, this is 
computationally demanding.  The LDA functionals in this paper include the EXX 
formalism, so the analysis and results here will be useful in the next development 
stage of DFT where EXX-compatible correlation is considered in detail.  In particular, 
since the interaction here is local, certain difficult-to-model long-range aspects of 
correlation will be isolated.  For example, contact-interacting fermions will still 
exhibit spin-density waves and long-ranged entanglement, two problems extremely 
difficult to model using traditional density functional methods.

The correlation energy functional is often modeled using the correlation energy of a solvable reference system of fermions.  For the chosen 1D interaction, this is the 
Gaudin-Yang model solved exactly via the Bethe-Ansatz technique 
\cite{G67,Y67,FB80}. The correlation energy per particle of the uniform system, 
$\epsilon\c^{unif.}$, can then be parameterized as in Refs. \cite{MB4} and 
\cite{XPAT6} to reproduce the exact curve. The high-density expansion is
\bea
\epsilon\c (n)= & & \nonumber\\
-\lambda^2/24 & + \lambda^3  \zeta(3)/(2\pi^4 n) - \lambda^4  0.00094/n^2 +{\cal O}(\lambda^5/n^3). & \nonumber\\
\label{echd}
\eea
$\zeta(3)$ is the Riemann zeta function evaluated at 3.  See appendix A for the calculation of the
second term.  

The low-density correlation energy is
\bea
\epsilon\c (n) = & & \nonumber \\
-\lambda n /4  & + n^2 \pi^2 / 8  -n^3 2\pi^2 \log(2) /(3\lambda) +{\cal O}(n^4/\lambda^2). & \nonumber\\
\label{ecld}
\eea   
The low density limit can be understood by noting that the interaction is so strong
that it mimics Fermi repulsion, so it must cancel the Hartree and exchange terms and add a
kinetic-like contribution to the energy. The third term is referred to in Ref. \cite{ABGP4,RFZ3}. 

A modified parameterization of the correlation energy per particle as a (4,4) Pad{\'e} is
\bea
\epsilon\c^{unif.}( n) =
\frac{A n^3 + B n^2 +C n}
{D n^3+E n^2 + F n + 1 }    
\nonumber\\ \label{epsclda}
\eea
with 
$A=-7.031 \; 951$,
$B=-2.169 \; 922$,
$C=-0.25$,
$D=168.766 \;  814$,
$E=77.069 \; 721$,
and
$F=13.614 \; 491$. 
This parameterization gives 3 terms in both the high- and low-density expansions of 
the correlation energy.  The error is less than 0.5$\%$ error in the 
correlation energy per particle for all densities.  We note the remarkable fact that 
parameters, $A$-$F$, are determined exactly with no approximate numerical fit.  
Details are given in the appendix B.

The LDA correlation energy functional is an integral over local contributions of the reference system's correlation energy per particle times the fermions per unit volume,
\bea
E\c^{LDA} [\n,\zeta] =
\int dx\; n(x) \;\epsilon\c^{unif.}( n(x)) f(\zeta(x)) . & & \label{eclda}
\eea
The exact high-density limit for $f(\zeta)$ can be obtained via diagrammatic 
perturbation theory (Appendix C) and is approximately $f(\zeta)\approx (1-\zeta^2 )$ 
used here for all densities.  The improved (4,4) Pad{\'e} parameterization of the LDA 
correlation energy functional is used because the ones given in Ref. \cite{MB4} and 
\cite{XPAT6} do not accurately reproduce the high density correlation kernel needed 
for TDDFT as we will explain later.

The DFT solution is obtained through a self-consistent solution of the Kohn-Sham 
equations using a modified version of the DFT code in Ref. \cite{MB4}.  In this code, 
a Numerov integration scheme is combined with the shooting method to obtain solutions of the Kohn-Sham equations.  
The number of grid points is chosen to converge energies to within mHartree accuracy, 
and the output is checked against an analytic EXX solution given in appendix D.

Time-dependent DFT allows the determination of the excited states.  The exact excitations occur at the poles of 
the density response-function \cite{RG84,C95}. Finding the excited-state transition 
energies amounts to the solution of a generalized eigenvalue problem:
\bea
\Omega_{ij\sigma,kl\tau}(\omega_{(I)})  F_{(I),kl\tau} = \omega_{(I)}^2 F_{(I),ij\sigma}
\eea
with 
\bea
\Omega_{ij\sigma,kl\tau} (\omega)
=  &
\delta_{\sigma\tau}\delta_{ik}\delta_{jl} 
\left(\epsilon_{j\tau}\!-\!\epsilon_{k\tau}\right)^2 & \nonumber\\
+ 2
K_{ij\sigma,kl\tau} (\omega)
&
\! \sqrt{(f_{i\sigma}\!\!-\!\!f_{j\sigma})
(\epsilon_{j\sigma}\!\!-\!\!\epsilon_{i\sigma})}
&
\sqrt{(f_{k\tau}\!\!-\!\!f_{l\tau})
(\epsilon_{l\tau}\!\!-\!\!\epsilon_{k\tau})} \nonumber\\
\label{tddft}
\eea
where 
$\epsilon_{i\sigma}$ is the KS eigenvalue of the $i$-th KS orbital of spin $\sigma$.  
$f_{i\sigma}$ is 1 if the $i$-th orbital of spin $\sigma$ is occupied; otherwise, $f_{i\sigma}$ is 0.  
$\omega_I$ is the $I$-th excitation value sought.
$K_{ij\sigma,kl\tau}(\omega)$ depends on the exchange correlation kernel as follows:
\bea
K_{ij\sigma,kl\tau} (\omega) = & &  \nonumber\\
 \int dx \;dx'
\phi_{i\sigma}^\ast(x)
\phi_{j\sigma}^\ast(x)\;
f_{\hxc,\sigma\tau}(x,x',\omega) \;
\phi_{k\tau}(x')
\phi_{l\tau}(x') . \nonumber\\
\eea
In general, the matrix $K(\omega)$ depends on the energy, $\omega$, of the solution, and this greatly complicates the
solution of the general problem.  However, the $\omega$ dependence is often ignored in what is
called the adiabatic approximation.

The adiabatic LDA kernel can be derived from the second functional derivative of the
time-independent LDA exchange correlation functional. Hence,
\bea
f_{HXC\;\sigma\tau}^{LDA}(x,x',\omega) 
= 
\frac{\delta^2 E\hxc^{LDA}[n,\zeta]}{\delta n_\sigma(x) \delta n_\tau(x')} 
= \nonumber \\
\lambda \delta(x-x')\left(1-\delta_{\sigma\tau}\right)  
+ \Big(\frac{d^2  [n(x) \epsilon_{C}(n(x)) ] }{d^2 n(x)}  +  2 
\frac{\epsilon_C(n(x))} {n(x) }
(1-2\delta_{\sigma\tau})  
\Big)\!
\delta(x-x') \!
\nonumber \\
= 
f_{HX\;\sigma\tau}^{EXX}(x,x') 
+ f_{C\;\sigma\tau}^{LDA}(x,x'). \nonumber \\
\label{fxc}
\eea
The kernel is split into two parts, an Hartree-exchange part ($f_{HX\;\sigma\tau}^{EXX}$) and
a correlation part ($ f_{C\;\sigma\tau}^{LDA}$).  The former is known exactly, the latter must be
approximated.  The result after the second equal sign is tailored specifically to the application in this manuscript assuming an unpolarized $\zeta=0$ system.  

Now, we will explain why we need to have an accurate correlation energy per-particle 
to third order in the high-density limit. The second derivative of the correlation 
energy density with respect to the density represents the heart of the adiabatic TDDFT 
approximation as it carries all the correlation effects beyond what is modeled in the 
KS orbitals.  The correlation kernel is the second functional derivative of Eq. 
\ref{eclda} with Eq. \ref{epsclda} plugged in explicitly. In the high-density limit, 
the first two terms of its second derivative vanish, and only the third and higher terms remain.  Therefore, in 
order to model the correlation kernel \emph{at all} in the high density limit, the 
third term must be included.  For this reason, we had to replace the (3,3) forms with 
the (4,4) Pad{\'e} given in this paper.  In 3D, the situation is different as the 
logarithmic dependence on the density means that the second derivative is divergent and non-vanishing; however, any additional finite terms might be neglected, and third 
order non-logarithmic terms might become important in certain common density ranges.

The \1dh2 model at hand has at most 2 bound KS orbitals, so we use the two-state 
single-pole approximation to TDDFT, first presented by Casida in Ref. \cite{C95}. It 
is assumed that the contribution of any finite number of unbound orbitals to the 
response function is negligible due to box normalization.  It is possible however, in 
the limit of extremely weakly bound orbitals, that this approximation is no longer 
valid. The solution of Eq. \ref{tddft} within the two-state model is \bea \omega_S = 
\sqrt{ (\epsilon_1-\epsilon_0) [ (\epsilon_1-\epsilon_0)+2 (K_{\uparrow\uparrow} + 
K_{\uparrow\downarrow})]} \eea for the singlet excitation and \bea \omega_T = \sqrt{ 
(\epsilon_1-\epsilon_0) [ (\epsilon_1-\epsilon_0)+2 (K_{\uparrow\uparrow} - 
K_{\uparrow\downarrow})]} \eea for the triplet excitation.  $\epsilon_0$ corresponds 
to the lowest-energy \emph{gerade} KS orbital eigenvalue and $\epsilon_1$ corresponds 
to the \emph{ungerade} excited KS orbital eigenvalue.  We use $\epsilon_0$ and 
$\epsilon_1$ here to distinguish these values from the exact non-interacting 
eigenvalues $\epsilon_+$ and $\epsilon_-$ even-though the KS states are also 
characterized as \emph{gerade} and \emph{ungerade}.  $K$ is a 2$\times$2 matrix in 
spin given by \bea K_{\sigma\tau}= & \lambda(1-\delta_{\sigma\tau}) 
\int_{-\infty}^\infty dx \phi_0^2(x) \phi_1^2(x) & \nonumber\\ \!+\! 
\int_{-\infty}^\infty \int_{-\infty}^\infty & dx\;dx'\; \phi_0(x) \phi_1(x) 
f_{C\;\sigma\tau}^{LDA}(x,x',\omega) \phi_1(x') \phi_2(x') . & \nonumber\\ \eea In 
section \ref{s:results}, we will directly compare the exact excited-state spectrum for 
Eq. \ref{1dh2} and the TDDFT spectrum.

DFT will likely fail for large separations, $a$, because the local treatment of 
correlation does not cancel the exchange and Hartree terms. This is the long-ranged 
self-correlation error, sometimes called the static correlation problem in DFT, and it 
exists even when the interaction is local.  The net effect is that two well separated 
fermions interact in the LDA while in a realistic system, the fermions would be 
entangled but otherwise non-interacting.  For ground-state DFT, we suggest a simple 
scheme to model the long-range correlation and to cancel the self-correlation error in 
the stretched case.  First, note that in the stretched situation, there is no problem 
in the polarized case since two fermions of the same spin do not interact via the 
contact interaction.  In the unpolarized case when $\zeta=0$, opposite spin particles 
are likely to localize in different regions of space, the interaction energy will be 
much less than expected if they delocalized.  In order for this to be captured, 
the correlation energy must exactly cancel the Hartree-exchange energy. This is 
achievable if $\epsilon_C(n) = -\lambda n / 4$ when $\zeta=0$, the negative of the 
Hartree and exchange energy.  The solution is then to obtain information from the 
density that fermions are in the \emph{stretched} regime and to apply long-ranged correlation in 
this case.  A previous attempt at describing this long-ranged correlation used the 
pair density function with some success \cite{PSB95}.  Another way this can be 
achieved is by relying on the KS orbitals and defining the dimensionless and unitary 
parameter, \bea \tau(x) = \frac{\sum_{occ.} |\phi_{i,\sigma}|^4(x) }{(\sum_{occ.} 
|\phi_{i,\sigma}|^2(x) )^2}, \label{localparam} \eea where $\phi_{i,\sigma}$ 
represents the $i$-th KS spin-orbital, and the sum is over $i$ and $\sigma$.  If the 
fermions are isolated as in the case of a one fermion system, $\tau(x)=1$; otherwise, 
$\tau(x)<1$.  For a two fermion system such as Diracium, $\tau(x)=1/2$, and 
$\tau(x)=0$ for the uniform reference system.  Physically, $\tau$ can be thought of as 
a measure of the inverse number of fermions that are locally relevant, and in this 
way, transcends the definition given here.  Unitary $\tau$ implies that a certain 
region of space is occupied by only one fermion and so many body effects are trivially 
unimportant.  As $\tau$ gets smaller in magnitude, the importance and nature of many 
body effects becomes important.

We restrict ourselves to the case $\zeta=0$. 
Suppose that we use $\tau(x)$ to model a local toggling between uniform-reference-system-based DFT
and the exact long-ranged limit:
\bea
E\C^{LC}[n] = & & \nonumber \\
\int_{-\infty}^\infty \!\!dx\; & \left[ f(\tau(x)) \epsilon_C^{LDA}(n(x),0) \!-\! (1\!-\!f(\tau(x)))  \epsilon_{HX}(n(x),0) \right] n(x) . & \label{toggle} \nonumber \\
\eea
$f(\tau)=0$ if $\tau=1$ and otherwise has a value of unity for the two fermion system, but in general, a more
complicated form is necessary.  
We call this functional
LC for \emph{locally cognizant} referring to the fact that the approximation is similar to the LDA
but allows for the recognition and proper treatment of single particle regions. Eq. \ref{toggle}
produces the same LDA results for Diracium and the small-spacing double-well problem since $\tau(x)=1/2$ everywhere for
these systems.  For stretched \1dh2, $\tau(x)=1$ for the exact solution.  LDA gives
$\tau(x)=1/2$ and the wrong energy. 

To handle the large separation limit, we need to consider details about the Kohn-Sham 
reference system. In this limit, the highest occupied and lowest unoccupied 
approximate KS orbitals are nearly degenerate, and it is plausible that the exact KS 
potential would result in orbitals that most closely resemble linear combinations of 
these.  If we were to express the exact KS orbitals in terms of the approximate ones, 
we would no longer be in a variational minimum of the approximate KS system, and 
consequently, the orbital energy contribution would rise.  Our approximate density 
function must then lower the interaction energy to compensate for this effect.

This LC functional will give the correct result if the interaction energy is lowered 
by a greater amount than the Aufbau rule raises the energy.  The scheme implies that 
symmetry can be broken. There are two lowest energy KS solutions: one with the up 
fermion localized right and the down localized left and vice verse.  The total density 
for each is the same, but the magnetization is inverted.  Since both solutions are of 
equal weight, the physical observable are expected to be averages of the equal energy 
states.  Thus, the total magnetization will vanish.  The philosophy here differs from 
the wave-function based idea of multiple determinants.

For example at $a=2$, the energy change of elevating the occupied orbitals is $\Delta 
T\s \approx 0.037$, and the change of turning on the long-ranged part, $\Delta E\c= 
E\c^{LDA}-E\hx=-0.157$.  These two conspire to lower the total energy.  For $a=1$, the 
$\Delta T\s=0.259$ dominates over $\Delta E\c=-0.195$, and the local correlation 
method is valid.  Examination of Eq. \ref{toggle} shows that LC will over-correlate 
relative to the LDA but will also increase the kinetic energy.  We will see later that 
this performance is required to improve upon the pure LDA.  An optimized effective 
potential scheme is needed to apply LC self-consistently, even to this 1D system and 
will be explored in future work.  The stretched limit poses challenges for TDDFT as 
well.  In the large separation limit, we should have $f_{C\;\sigma,\tau}(x,x',\omega) 
= -\lambda \delta(x-x') \left(1-\delta_{\sigma\tau}\right)$ because in the large 
separation limit $E\c=-E\hx$.  The form of Eq. \ref{fxc} is compatible with this limit 
if $\epsilon_C = - \lambda n/4$ as the exact correlation functional should be in this 
limit.  It would be interesting to known whether the self-interaction corrected LDA 
would correctly give this limit.  This too requires an optimized effective potential 
scheme.


\section{Method of Exact Solution}
\label{s:exact}

Here, we present a technique to exactly solve the eigenvalue problem, $\hat{H}\Psi(x_1,x_2)=E\Psi(x_1,x_2)$ using $\hat{H}$ in equation \ref{1dh2}.  
This work extends an idea originally introduced in Rosenthal's work 
\cite{R71} but involves many more challenges than the Diracium solution.

The exact spin-singlet real-space wave-function of two 1D fermions is a 2D function, 
$\psi(x,y)$, constrained by particle interchange rules to be either symmetric or 
antisymmetric under the swapping of $x_1$ and $x_2$.  The eigenvalue problem is difficult to 
solve using the traditional wave-function based methods of quantum chemistry such as 
configuration-interaction, because the virtual spectrum is mostly unbound. A 
prohibitively large number of excited configurations would be needed to provide an 
accurate solution.  Perturbative approaches suffer similar limitations. A variational 
method could provide a highly accurate solution.  However, the solution would be 
affected by the assumed form of the variational wave-function. This form, in 
principle, limits the accuracy of the many-body solution.  In this section, we present 
an exact numeric solution.  By exact numeric, we mean an integral equation that can be 
solved to arbitrary accuracy numerically by increasing and refining the number of integration points.

The exact solution is found by reducing the Schr{\"o}dinger equation with the Hamiltonian in Eq. \ref{1dh2} to a set of coupled integral
equations.  To do this, the differential equation is expressed in momentum space in terms of three 1D trace
functions:
\bea
G_1(k) = \int_{-\infty}^\infty dy\; \exp(iky) \left[\psi(a,y)+\psi(-a,y)\right],
\label{G1}
\eea
\bea
G_2(k) = i\int_{-\infty}^\infty dy\; \exp(iky) \left[\psi(a,y)-\psi(-a,y)\right],
\label{G2}
\eea
and
\bea
H(k) = \int_{-\infty}^\infty dy\; \exp(-iky) \psi(y,y).
\eea
Note that we use $y$ here as the conjugate spatial coordinate to $k$ to avoid confusion with $x_1$ and $x_2$. 
The $k$-space wave-function solution can be expressed algebraically in terms of these trace functions:
\begin{eqnarray}
\Phi(k_1,k_2) = & & \nonumber \\
\frac{2}{k_1^2+k_2^2+p^2 } &
(
\cos(k_1a) G_1(k_2)
+\cos(k_2a) G_1(k_1) &  \nonumber\\
+\sin(k_1a) G_2(k_2) &
+\sin(k_2a) G_2(k_1)
-\lambda H(k_1+k_2) ) . &  \nonumber\\
\label{pspace}
\end{eqnarray}
The many-body energy eigenvalue is $E=-p^2/2$ defining $p$.

To obtain a set of 1D integral equations, we Fourier transform Eq. \ref{pspace} 
into real space and use the result to express $\psi(a,y)$ and $\psi(-a,y)$. Then, we 
plug $\psi(a,y)$ and $\psi(-a,y)$ into the right-hand sides of Eqs. \ref{G1} and 
\ref{G2}.

For $S_0$, the ground-state, we write out the resulting coupled integral equations for $G_1$ and $G_2$ explicitly,
\bea
G_1(a k) =
\frac{2}{\pi} \left(
1-a
\frac{(1+\exp^{-2\sqrt{(a k)^2+(a p)^2} } )}
{\sqrt{(a k)^2+(a p)^2}  }
\right)^{-1} \nonumber \\
\int_0^\infty\!\!\!\!\!\! d(a k')
\Big[
\frac{2a \cos(a k) \cos(a k')}
{ (a k)^2 +(a k')^2 +(a p)^2}
\nonumber \\
- \frac{2\lambda a^2}{\pi} 
(\kappa_{cc,1,a p, a \lambda} (-a k, -a k')
+ \kappa_{cc,1,a p, a \lambda} (-a k, a k'))
\Big] \! G_1(a k') \nonumber \\
- \frac{2\lambda a^2}{\pi} 
(\kappa_{cs,1,a p, a \lambda} (-a k, -a k')
-\kappa_{cs,1,a p, a \lambda} (-a k, a k'))
\! G_2(a k')  \nonumber \\
\label{g1}
\eea
and
\bea
G_2(a k) =
\frac{2}{\pi} \left(
1-a
\frac{(1+\exp^{-2\sqrt{(a k)^2+(a p)^2} } )}
{\sqrt{(a k)^2+(a p)^2}  }
\right)^{-1} \nonumber \\
\int_0^\infty\!\!\!\!\!\! d(a k')
\Big[
\frac{2a \sin(a k) \sin(a k')}
{ (a k)^2 +(a k')^2 +(a p)^2}
\nonumber \\
 \!-\!
 \frac{2\lambda a^2}{\pi} 
(
\kappa_{ss,1,a p, a \lambda} (-a k, -a k')
-\kappa_{ss,1,a p, a \lambda} (-a k, a k'))
\Big] \! G_2(a k') \nonumber \\
- \frac{2\lambda a^2}{\pi} 
(\kappa_{sc,1,a p, a \lambda} (-a k, -a k')
+\kappa_{sc,1,a p, a \lambda} (-a k, a k'))
\! G_1(a k')  \nonumber \\
\label{g2}
\eea
with
\bea
 \kappa_{cc,a,p,\lambda} (k, k') = & & \nonumber \\
& \int_0^\infty dq
\Big(1+\frac{\lambda}{\sqrt{q^2+2p^2}}\Big)^{-1} &\nonumber \\
\Big[
 \Big(\frac{\cos[a(k +q)]}{(k \!+\! q)^2+k^2+p^2}\Big)
 \Big(\frac{\cos[a(k'+q)]}{(k'\!+\! q)^2+k'^2+p^2}\Big) &\nonumber\\
+\Big(\frac{\cos[a(k -q)]}{(k \!-\! q)^2+k^2+p^2}\Big)
 \Big(\frac{\cos[a(k'-q)]}{(k'\!-\! q)^2+k'^2+p^2}\Big) \Big]
. & \label{kappa} \nonumber \\
\eea
We have placed pre-factors of $a$ in locations that are convenient for numerical reasons.  A similar set of 
equations can be written down for the $S_1$ state.  $S_2$ satisfies the above set with a
different $p$. $\kappa_{cc,a,p,\lambda}(k_1,k_2)$ can be evaluated explicitly using complex
analysis.  The notation $cc$ stands for the trigonometric functions that are contained in $\kappa$. 
$cc$ is for $\cos\cos$, $cs$ for $\cos\sin$, and so on.  The square-root introduces a branch cut affecting the contour integration that is 
best handled through Gauss-Legendre numerical integration.

So, we arrive at a set of coupled 1D integral equations in two functions $G_1(k)$ and
$G_2(k)$. Converting the integrals to Gauss-Legendre sums allows us to express the coupled integral
equations as a matrix problem:
\bea
\left( \begin{array}{c}
G_1\\
G_2\end{array}
\right)
= \mu(p)
\left( \begin{array}{cc}
M_{11} & M_{21}  \\
M_{12} & M_{22} \\
\end{array}
\right)
\left( \begin{array}{c}
G_1\\
G_2
\end{array}
\right)
\eea
where $\mu(p)$ is an eigenvalue that equals unity when the appropriate $p$, that satisfies the
original eigenvalue problem is inputted.

In expressing the integral equation in matrix form, the integrals are discretized onto a set of
points.  There is no unique way to do this.  We chose a Gauss-Legendre inspired integration scheme. 
However, the implementation is not straightforward as the integrals take on the form,
\bea
\int_{-\infty}^\infty dk\; \frac{ \mbox{trig}^2 k }{(k^2+p^2)(k^2+p^2/2)} \xi(k),
\eea
where the function $\mbox{trig}^2 k$ can be $\cos^2 k $, $\sin^2 k$, or $\sin k \cos k$ and $\xi(k)$ is a smooth non-oscillatory function of k.  
To obtain accurate numerical quadratures, we break up the domain 
into intervals of $\pi/2$ and integrate each separately using a suitable method.  
For convergence to better than 10 nano-Hart. in energy and $10^{-8}$ in $\mu(p)$ at $a=1$, we need
800 grid points for the primary integration region and 200 additional tail points to model a portion
of the asymptotic tail.  The complicated scheme is highly
accurate and has been verified by producing the exactly known non-interacting results to 9
significant figures for 1000 k-points.

Two tests verify the accuracy of the exact solution.  First, the exact solution is 
shown to approach the known united-atom limit. The idea is that this exact solution 
should give results approaching Diracium's with an appropriate combined potential 
strength. For two wells with $Z=1$ each, we get Diracium with $Z=2$. Figure 
\ref{f:etotal} shows the energy for \1dh2 for various $a$ and methods, and the 
approach can be seen.  The approximate methods will be discussed in section 
\ref{s:results}.  The far left limit on the plot is Diracium when $\lambda=1$ and 
$Z=2$, $E_2=-3.155$. For crushed $H_2$, we find $E=-3.023$ when $a=0.01$ and 
$E=-3.087$ when $a=0.005$ extrapolating to $E_2=-3.152$ in excellent agreement with 
the united-atom limit.

As a second test, the exact numeric solutions are shown to satisfy the generalized 
virial relationship given by Eq. \ref{virial} to within 10 mHart.  The accuracy is modest 
due to the numeric challenge of solving the required triple integrals with the limited 
sampling of $k$ points.  For $Z=1$ and $\lambda=1$, the left-hand side of Eq. 
\ref{virial} is $-318$ $\mu$Hart. while the right-hand side is $-317$ $\mu$Hart. in less than 
perfect but still acceptable agreement.


\section{Comparison of TDDFT and Hubbard models to the exact results}
\label{s:results}

In this section, we compare adiabatic TDDFT and the Hubbard model to the exact 
results.  To start, we make some comments about the parameter regimes chosen.  This 
model offers a rich spectrum of phenomena; however, for the sake of brevity, we have 
restricted ourself to a rather arbitrarily chosen parameter ranges.  As mentioned, 
we set $Z=1$ in this section for convenience, but results for arbitrary $Z$ can be 
related via scaling.

We analyze the energy spectrum of \1dh2 versus $\lambda$ for two values of 
$a$.  The first case is the double-well potential with 2 bound-fermions and separation 
$a=1$.  In this case, we expect DFT to provide an accurate description of the spectrum 
since the fermions are both localized in the area of the double well.  The Hubbard 
ground-state is expected to be too low by at least $68$ mHart, the amount that the 
kinetic energy is misrepresented in the non-interacting case.  For the second case 
$a=2$, the system is in the stretched H$_2$ limit.  The LDA is known to be unreliable 
in this limit because its failure to capture the localization of fermions to opposite 
sites without symmetry breaking.  On the other hand, the Hubbard model is designed to 
work well in the stretched case.  For example, the non-interacting Hubbard error at 
$a=2$ is less than 3 mHart. for the ground-state, $S_0$.  For now, our primary focus 
is on comparing stable results, and thus, we restrict our analysis to $\lambda<2$, 
safely within the bound regime for the ground-state, $S_0$.

\begin{table} 
\begin{center}
\begin{tabular}{l|cc|cc|c}\hline
                     & nonint. occ. orb. & &  exchange  & symmetry    &  two-body  \\
State                &  1   &  2       & space & spin  &  energy \\ \hline
$S_0$ & +  &  +   & Symm. & Asymm.&  $2\epsilon_+$ \\

$S_1$ & +  &  -   & Symm. & Asymm.&  $\epsilon_+ + \epsilon_-$  \\  
$S_2$ & -  &  -   & Symm. & Asymm.&  $2\epsilon_-$ \\
$T_1$ & +  &  -   & Asymm.& Symm. &  $\epsilon_+ + \epsilon_-$ \\ \hline
\end{tabular}
\caption{\label{t:spec} The non-interacting many-particle bound-states of the Hamiltonian Eq.
\ref{1dh2}.  $+$ stands for the \emph{gerade} single particle orbital with eigenvalue $\epsilon_+$,
and $-$ stands for the \emph{ungerade} single particle orbital with eigenvalue $\epsilon_-$.  }
\end{center}
\end{table} 

Our labeling scheme for the quantum states is motivated by the noninteracting 
many-particle spectrum.  The non-interacting multiple particle spectrum can be 
constructed from the single particle states. For the 2-particle solution, products of 
the single particle orbitals must be properly symmetrized. The entire bound spectrum 
of the non-interacting 2 particle states is given in Table \ref{t:spec}. Note that the 
non-standard labeling of the states is used because angular momentum needed for the 
traditional labeling scheme is not well-defined in 1D. There are 3 spin-singlet states 
and 3 spin-triplet states.  The spin-singlet states have symmetric wave-functions 
under particle interchange.  There are 2 single-particle orbitals, \emph{gerade} and 
\emph{ungerade} ones, and 3 unique symmetric products can be made of these.  The 
triplet state is triply degenerate in the spin manifold but has only one spatial 
contribution, an anti-symmetrized product of the \emph{gerade} and \emph{ungerade} 
orbitals.  Antisymmetric products of like noninteracting orbitals vanish.  It is 
pointed out that the singlet-state comprised of a product of 2 \emph{ungerade} 
orbitals represents a double excitation from the ground-state.  The states are labeled 
$S_0$, $S_1$, and $S_2$ in the singlet manifold, and $T_1$ in the triply degenerate 
triplet manifold.  A transition form $S_0\rightarrow S_1$ is the first singlet 
excitation, and a transition form $S_0\rightarrow T_1$ is the first triplet 
excitation.  Both are calculable in first-order linear response theory of DFT.  A 
transition form $S_0\rightarrow S_2$ is a double excitation, proportional to the 
intensity of light squared, and is formally beyond first order response theory.  This 
fact can be realized by considering the noninteracting response function and noting 
that no double poles exist.  Nevertheless, the bare Kohn-Sham (KS) double excitations might be 
considered a good first approximation to the double excitation due to the linear response terms 
vanishing and a better approximation than the KS 
single excitations were to their counterparts.

The interacting two-fermion spectrum is expected to have a one-one correspondence with 
the noninteracting one excepting the possible disappearance of the highest energy 
states into the continuum as the interaction, $\lambda$, is increased.  Take for example, 
the 1D analog of Helium, Diracium, at $Z=1$.  A single $\delta$-well typically can 
bind two fermions of opposite spin in 1D, but above $\lambda_{crit.} = 2.6673532258$, 
the two-particle state merges with the continuum, and only one fermion can be bound 
\cite{R71,CDR6}.  For 
two-fermions in a double-well, the critical interaction strength, $\lambda_{crit.H_2}(a,S)$, depends on the well-spacing, $a$, and state, $S$.  The 
critical interaction strength, $\lambda_{crit.H_2}(a,S)$, is likely larger than its 
corresponding value for a single well due to the stabilization effects of the 
hybridized orbitals or, in Hubbard parlance, the energetic favor-ability of hopping.  
Most likely, each excited state in the double well has a different critical value as 
the higher energy states are likely to vanish at smaller $\lambda$ than the ground-state.  However, the 
triplet state is interaction independent because like spins do not experience the 
contact interaction.


\begin{figure}[t]
\centering
\includegraphics[width=2.9in]{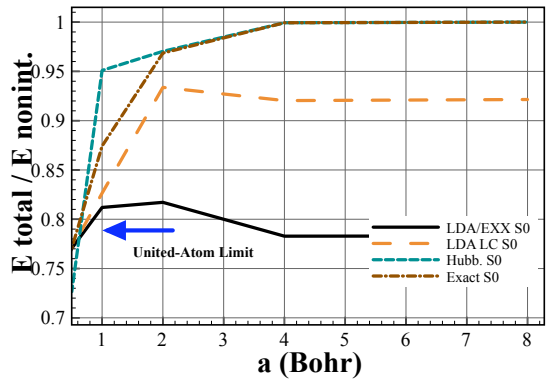}
\caption{(Color online) Normalized ground-state energies for the 1D analog of H$_2$ at separation $a$ and interaction strength, $\lambda=1$, within several approximations.  The results are normalized by the noninteracting result.  In the asymptotic limit, the ratio should be unity.  The solid line is the LDA result, the dotted line is the LDA LC result, the medium-dashed line is the Hubbard result, the long-dashed line with diamonds is the exact result, and the arrow indicates the united-atom Diracium limit.}
\label{f:etotal}
\end{figure}


\begin{figure}[t]
\centering
\includegraphics[width=2.9in]{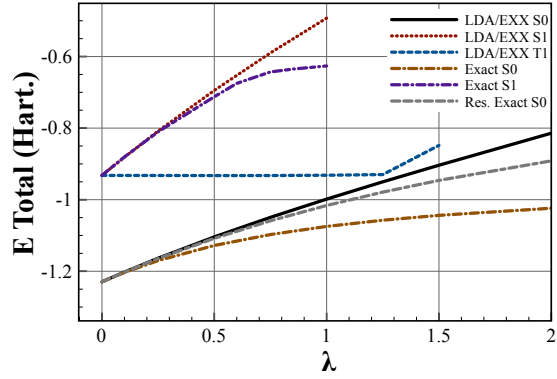}
\caption{(Color online) DFT and exact energy spectrum in atomic units for the 1D analog of H$_2$ for various $\lambda$ with inter-atomic separation, $a=1$. These results are generated through the
exact solution of the quantum many-body problem and the LDA version of TDDFT.  $S_0$ and $S_1$ are the ground-state and first singlet excited state, and $T_1$
is the first triplet excited state.  The solid line is the LDA $S_0$ result, the gray varyingly dashed line is a restricted version of the exact result, 
 the long-dashed line with diamonds is the exact result for $S_0$,  the dotted line is the
LDA $S_1$ result, the medium dashed line is LDA $T_1$ result, and the alternating short-dashed line is the exact $S_1$
result.}
\label{f:e1a} 
\end{figure} 


\begin{figure}[t]
\centering
\includegraphics[width=2.9in]{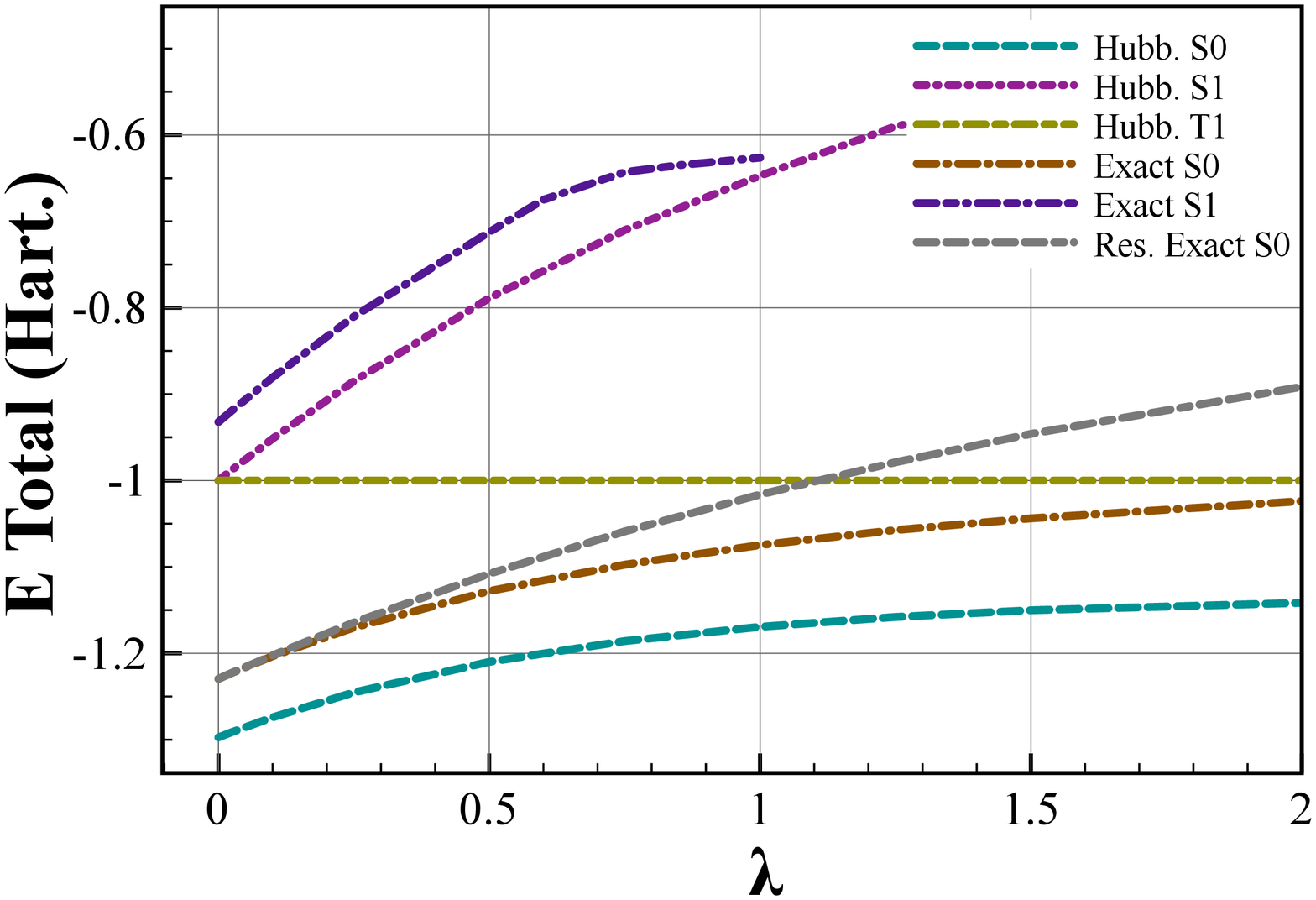}
\caption{(Color online) Hubbard and exact energy spectrum in atomic units for the 1D analog of H$_2$ for various $\lambda$ with inter-atomic separation, $a=1$. 
The long-dashed line with diamonds is the exact result for $S_0$,   the alternating short-dashed line is the exact $S_1$
result,
 the long-dashed line is the Hubbard $S_0$, the alternating short-long line is the Hubbard $S_1$ result,
 and
 the alternating medium-dashed line is the Hubbard $T_1$ result. }
\label{f:e1b} 
\end{figure} 


\begin{figure}[t]
\centering
\includegraphics[width=2.9in]{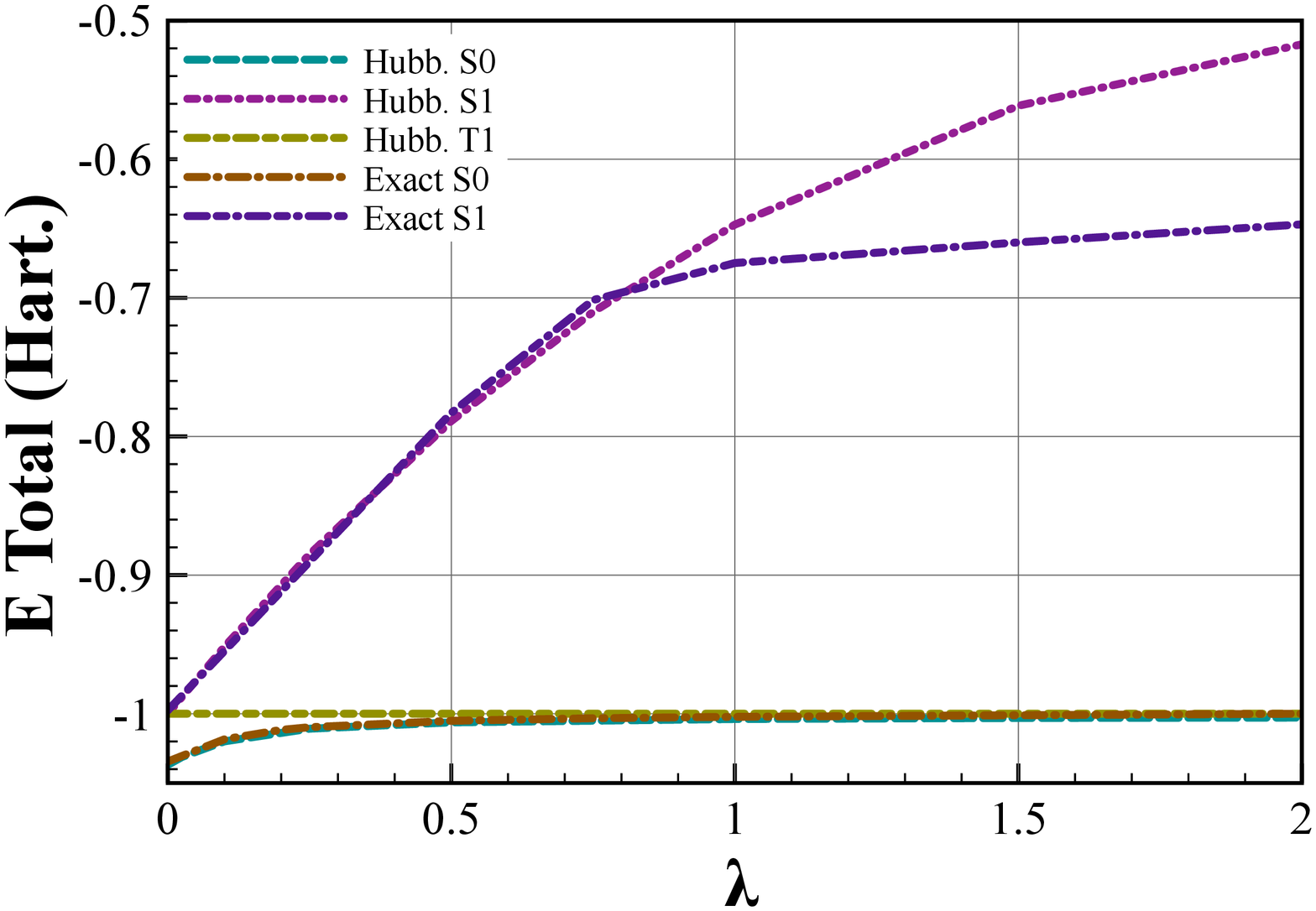}
\caption{(Color online) 
Hubbard and exact energy spectrum in atomic units for the 1D analog of H$_2$ for various $\lambda$ with inter-atomic separation, $a=2$.   The long-dashed line with diamonds is the exact result for $S_0$,  the alternating short-dashed line is the exact $S_1$
result.,
the alternating medium-dashed line is the Hubbard $S_0$, and finally, the alternating long-dashed line is the Hubbard $S_1$ result.}
\label{f:e2b} 
\end{figure} 


\begin{figure}[t]
\centering
\includegraphics[width=2.9in]{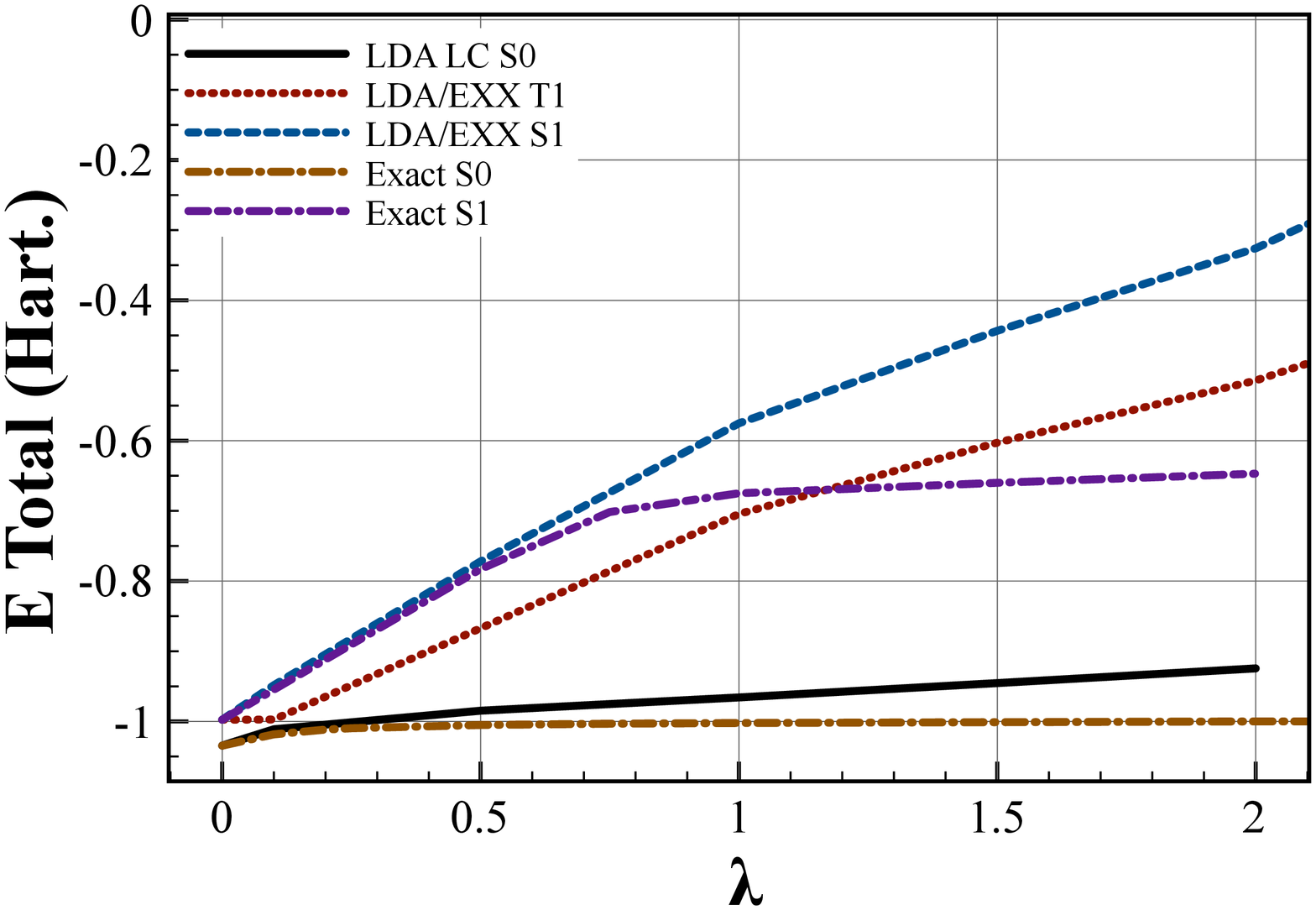}
\caption{(Color online) 
DFT and exact energy spectrum in atomic units for the 1D analog of H$_2$  for various $\lambda$ with inter-atomic separation, $a=2$. 
The long-dashed line with diamonds is the exact result for $S_0$, 
 the alternating short-dashed line is the exact $S_1$
result,
the dotted line is LDA LC $S_0$ result, the medium-dashed line is the DFT $T_1$ result, and the long-dashed line is the DFT $S_1$ result.}
\label{f:e2a} 
\end{figure} 


In Fig. \ref{f:etotal}, the total energies are normalized by the non-interacting 
results for $\lambda=1$.  In the limit of large $a$, the 
ratio should become unity as the well-separated fermions will not interact but will 
reside on different sites. The exact solution is globally spin unpolarized but locally 
acquires a nontrivial spin dependence.  This is the statement that the two fermions 
will occupy different sites and break symmetry.  The exact $S_0$ result here is the 
fully non-constrained solution of the \1dh2 Hamiltonian.  The Hubbard model $S_0$ 
reproduces the long range charge separated limit accurately but fails to describe the 
crushed limit due to improper scaling of the hoping term in this limit.  This is 
easily seem through the non-interacting scaling behavior of the hoping term.  The 
LDA/EXX $S_0$ does approach the crushed atom limit as is expected from earlier work on 
the Diracium system.  However, the LDA alone fails to capture long-ranged charge 
separation and fails to capture about 25\% of the total energy in the stretched limit.  
A perturbative application of the LDA LC $S_0$ greatly improves this approach since it 
allows for opposite spins to become separated.  In the intermediate range, $a\sim 
1-2$, the DFT methods and Hubbard model bracket the exact result indicating that both 
proper scaling and ability to localize particles are vital in this range.

In Fig. \ref{f:e1a}, we see the energy spectrum plotted for \1dh2 with $a=1$ at 
various interaction strengths, $\lambda$.  The exact result does not level out with 
increasing interaction strength. The system probably ionizes at some critical 
interaction strength as was the case for Diracium.  Exact ground-state (Exact $S_0$) and 
restricted exact ground-state (Res. Exact $S_0$) results are presented.  In the later, the 
solution is forced to be expressed purely in terms of $G_1(k)$ with the antisymmetric 
$G_2(k)$ forced to vanish (See Eq. \ref{g1} and \ref{g2}) .  In wave-function theory, 
this corresponds to unbroken symmetry.  The result is a higher energy solution than 
the true broken symmetry ground-state.  The difference for weak interactions is 
negligible but at larger interactions the restriction causes an energetic error 
exceeding 10\%, and a qualitative prediction of the cross over between the triplet 
excited-state and the ground-state.  The exact result does not cross over.  This 
finding highlights the importance of symmetry breaking in describing static 
correlation in the design of density functionals.

LDA/EXX $S_0$ and the restricted exact $S_0$ results are in excellent agreement until 
$\lambda=1$.  Beyond this point the two deviate slightly but increasingly.  This is 
most likely due to the difficulties of approaching the ionization threshold where the 
density is less localized.  The $S_1$ states agree up to $\lambda=0.5$.  The LDA $S_1$ 
fails to display the leveling off of the exact $S_1$ result.  The DFT triplet (LDA/EXX 
$T_1$) result like the Hubbard (shown in Fig. \ref{f:e1b}) and exact results is 
constant.  However, at $\lambda=1.2$, the triplet state becomes unreliable.  This 
is due to the cross over between the DFT $\epsilon_0$ energy and the $\epsilon_1$ 
orbital energy.  The single pole approximation becomes numerically unstable and thus 
unreliable.

For the ground-state, KS theory performs quite outstandingly reproducing 98 \% of the 
restricted ground-state energy at $\lambda=0.5$.  In the 
strongly-interacting regime at $\lambda=2$, DFT still gives a result within 80 \% of 
the exact value.  EXX on the other hand would be more inaccurate giving about 60\% 
of the total energy.  The EXX method is related to the Hartree-only theory and
the nonlinear Schr{\"o}dinger equation that is popular in many treatments of 1D 
systems.  We see here an example that non-linear Schr{\"o}dinger approach does not work accurately or reliably for 
moderate to strong interaction strengths.  Although the DFT results perform exceptionally well 
up to moderate interaction strengths, the results are under-correlated for stronger 
interactions, $\lambda > 0.5$.  This is because partial localization of the fermions 
to opposite ends of the system is not accounted for.  The exact restricted results 
correspond to a solution of the exact integral equations with $G_2(k)$ forced to 
vanish.  This is valid for the non-interacting solution because in that case the 
eigenstates are single Slater determinants of non-interacting single-particle 
solutions.  For the interacting state, the \emph{restricted} result does not represent a true eigenstate of the original 
Hamiltonian and according to the variaitonal principle, has a higher energy than the exact ground-state. 
Notwithstanding, the \emph{restricted} energies agree quite well with the pure LDA values 
indicating that LDA correlation is adequate to describe the system if there were no 
localization.

TDDFT allows us to find the spectrum of excited-states.  For the $S_1$ state, we show 
only up to $\lambda=1$ in Fig. \ref{f:e1a}.  At larger interaction strengths, this 
state tends to be unstable and decays into an unbound Fermion and one bound Fermion.  As a 
measure of this accuracy for TDDFT in the two state model, we work backward.  The 
triplet state does not experience the interaction so the triplet excitation is known 
exactly without using the TDDFT formalism.  The triplet is unaffected by the 
interaction and should be a straight horizontal line with respect to $\lambda$.  Thus, 
TDDFT should reproduce this line if the kernel and orbitals are both accurate.  This 
is what is seen up to about $\lambda\approx 1.2$.  At this point the triplet energy 
and the singlet restricted ground-state energy are close hinting at a level crossing.  
The occurrence of a cross-over differs from the Hubbard model where no level crossing 
occurs.  The total energy excited state gap for the singlet is much larger than for 
the triplet gap.  It is interesting to note that the singlet gap at $\lambda=1$ is 
only about one quarter larger than at $\lambda=0$.  Adiabatic effects are important, 
but the ratio of the singlet to triplet gaps grows more than the energy of the singlet 
state.  The actual non-adiabaticity must be a very complicated functional of the 
excited-state energies to capture this behavior.  For the triplet to be well 
reproduced implies that the potential, orbitals, and kernel (at this energy range) are 
accurate.  But the prediction of the first singlet excitation is not as accurate.  
Since the orbitals and potential are the same for the singlet, the approximation of the kernel must 
not be as accurate in the calculation of the singlet.

The Hubbard $S_0$ is typically over-correlated as shown in Fig. \ref{f:e1b}.  This is, 
in part, due to overemphasis of the kinetic energy at this smaller spacing.  But the 
model is also qualitatively wrong for larger interactions as the energy goes to a 
constant while the exact result continues to grow with the interaction strength until 
the system is eventually ionized.  The over-correlation is due to improper scaling 
towards the crushed H$_2$ limit. Additionally, the two-site Hubbard Hamiltonian does 
not allow for unbound-states and is consequently incapable of describing ionization. 
In Fig. \ref{f:e1b}, the ground-state ($S_0$) Hubbard curve resembles the exact curve 
except for a offset of about 0.1 Hartree.  This lower energy is a result of the 
improper scaling of the kinetic energy hopping term in the united-atom limit.  The 
Hubbard $S_1$ state suffers from a similar offset problem, however, the $S_1$ state 
does not demonstrate the sharp leveling off of the exact result at $\lambda=0.6$.    The higher energy Hubbard result can be rationalized by realizing that for large $U$, there will be significant projection of the localized solutions onto the other site thus increasing the effective overlap.   This increased delocalization is not described by the Hubbard model.  The $T_1$ Hubbard curve is flat by construction since it does not depend on  $U$, the inter-particle interaction.

For the stretched case, we expect the Hubbard model to be essentially exact because 
the hopping term and single site repulsion terms are accurate.  This is in fact what 
in seen in Figure \ref{f:e2b}.  The remarkable agreement is not surprising as the 
single orbital overlaps decay exponentially, and the relevant hopping parameters are 
small.  The $T_1$ Hubbard curve is again level as explained in the previous plot.  For 
larger $\lambda$, the Hubbard result for $S_1$ is dangerously close to ionization 
$\approx -0.5$.  It was seen that local correlations are inadequate to properly 
describe ionization.

In Figure \ref{f:e2a}, the comparison between DFT and exact is less satisfactory.  The 
ground-state results deviate quite substantially for interactions just larger than 
$\;\lambda=0.1$.  The ground-state energy is in error due to the failure of LDA to 
account for the localization of fermions to opposite sites.  The LDA LC, applied 
perturbatively, does provide some of this information and thus drastically improves 
the agreement with the exact $S_0$ result.  A self-consistent application of this 
functional would involve an optimized effective potential algorithm that is beyond the 
scope of this work.  However, self-consistency is likely to improve the accuracy.  It 
would be interesting to find out how a kernel based on this improved functional would 
perform.  The perturbative approach only approximates the stretched $H_2$ limit while 
a fully self-consistent approach should exactly match at large $\lambda$.  LDA/EXX 
$T_1$ only agrees up to to $\lambda=0.1$.  The orbitals are not faithful 
representations of the exact KS orbitals since the LDA ground-state is inaccurate.  
Curiously, the LDA/EXX $S_1$ and exact $S_1$ agree exceptionally well up to 
$\lambda=1/2$.  This is because the local kernel cancels the self-interaction 
correlation error in the ground-state calculation.  No excited state results are reported 
for LDA LC TDDFT because the method had been applied perturbatively, and the 
self-consistent wave-functions and kernel are not available in the analysis.

\section{conclusion}
\label{s:conclusions}

In this paper, we have explored how two standard viewpoints of condensed matter physics 
describe an interacting 1D many-particle system.  One method, LDA DFT, 
provides accurate energies within about 0.1 mHartree for intermediate interaction 
strengths ($\lambda\leq0.5$) and distances ($a<2$) but fails appreciably at large well 
spacings. Analysis of the exact result shows that this limitation is due to symmetries 
induced by the short-sightedness of the restricted KS scheme within the LDA.  When the 
restricted symmetry is enforced on the exact solution, the result lies much more 
closely to the LDA indicating that the lack of localization is the key deficiency in 
the  restricted LDA KS treatment. To overcome this challenge, a local parameter, 
$\tau(x)$, is introduced that describes the effective local number of fermions and is 
readily implementable in existing electronic structure codes.  This local measure 
allows the use of two reference systems in the construction of a density functional: 
the uniform reference system and the single particle system. The introduced 
functional, when applied perturbatively, is shown to better reproduce the energy curve 
of \1dh2 versus well spacing.  A self-consistent application will require an optimized 
effective potential approach beyond the scope of this work but is likely to improve the 
agreement.

On the other hand, the two-site Hubbard model provides a qualitatively accurate 
description of the ground-state across a wide range of parameters describing both 
united-atom and separate-atom limits, but it fails in its quantitative predictions and 
has questionable scaling characteristics.  This is not surprising as the model is 
limited by design.  The 2-site Hubbard model does not include the continuum and will 
therefore fail to describe ionization and scattering.  

The excited-state results follow a similar pattern as the ground-state results.  DFT 
is accurate for small well spacings, and the Hubbard model is more reliable for larger 
spacings.  In TDDFT, the approximation of the higher energy singlet fails for weaker 
interactions than the lower energy triplet.  For the triplet to be well reproduced 
implies that the potential, orbitals, and kernel (at this energy range) are accurate.  
Since the first two are the same for the singlet, the approximation of the kernel must 
not be as accurate in the calculation of the singlet. A strongly non-adiabatic kernel 
would explain why TDDFT predicts the lower energy triplet but not the higher energy 
singlet.  Thus, non-adiabaticity outweighs ultra-nonlocality problems.  Perhaps, this 
is due to the locality of exchange for contact interactions and could be quite 
different from what occurs with long-ranged 3D interactions.   Curiously, the Hubbard first 
excited-state can also prove drastically wrong in cases when the fermions are strongly 
interacting, large $U$, as the Hubbard treatment forces same site localization while the exact system will have significant delocalization.   
This result has significant implications for the reliability of LDA+U results with large $U$.

The realm of 1D contact-interacting fermions provides an interesting opportunity to 
compare the Hubbard and DFT models in detail.  Results found here have provided 
insight into models of 3D Coulomb interacting systems where many of the underlying 
quantum many-body effects are obfuscated by the long-ranged nature of the interaction. 



\appendix
  \renewcommand{\theequation}{A-\arabic{equation}}
  \setcounter{equation}{0}  
  \section*{APPENDIX A}  


%
\begin{figure}[t]
\centering
\includegraphics[width=3.3in]{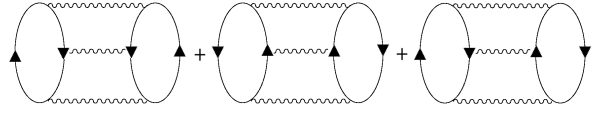}
\caption{
Diagrammatic representation of the third order contributions to the interaction energy.
Spin labels are omitted since the two loops in each diagram must have opposite spins.  Up arrows represent particles, and down arrows represent holes.  The final diagram contributes twice because a horizontal rotation produces a new diagram.}
\label{f:feyn1} 
\end{figure}

%
\begin{figure}[t]
\centering
\includegraphics[width=2.2in]{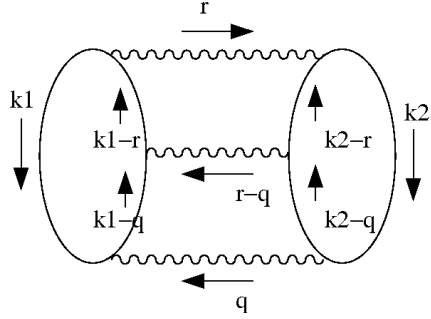}
\caption{
Momentum labels for the third order contributions to the interaction energy. 
$q$ and $r$ are momentum transfers.
$k_1$ and $k_2$ can label particle or hole momenta depending on the diagram.
}
\label{f:feyn2} 
\end{figure}

In this appendix, we  find the third-order $\lambda$ term in the  high density limit  of the correlation  energy per
particle for  Deltium, the one-dimension uniform fermion system, using  the Goldstone diagrammatic
approach to perturbation theory in  momentum space \cite{G57}.  

The Fourier  transform of the
interaction potential  is 
$V(q)  = \frac{\lambda}{L} \int_{-L}^{L} dx\; \delta(x) e^{i q x}  = \frac{\lambda}{L} $ 
where $L$ is the arbitrary length between the boundaries confining the system.
Like  spin fermions  do not  interact via  the  $\delta$-function, this
means that only vertices that  connect opposite spins enter into
the  diagrammatic series.   This  is a tremendous simplification as  many
diagrams vanish.  A further  simplification is that the interaction
is independent of the momentum transfer, $q$.

To  third order,  three different  diagrams contribute as seen in Fig. \ref{f:feyn1}.   
The momenta in the each diagrams are labeled according to the arrows in the 
two-bubble  diagram shown  in Figure  \ref{f:feyn2}.  From  the
standard rules  of perturbation theory, the third order term can be written as a sum of multi-dimensional integrals,
\bea 
N  \epsilon_C^{(3)} = 
\frac{\lambda^3}{L^3}  \frac{L^4}{16\pi^4} \frac{n_s}{2} 
\nonumber \\  
\Big(
\int_{-\infty}^\infty \!\!\!\!\!\! dq 
\int_{-\infty}^\infty \!\!\!\!\!\! dr 
\int_{k_F}^{\infty} \!\!\!\!\! d |k_1|  
\int_{k_F}^{\infty} \!\!\!\!\! d |k_2| 
\frac{1}{ q ( q + k_1-k_2) r (r + k_1-k_2) }  \nonumber \\
+\int_{-\infty}^\infty  \!\!\!\!\!\! dq 
\int_{-\infty}^\infty \!\!\!\!\!\! dr 
\int_{-k_F}^{k_F}  \!\!\!\!\! dk_1  
\int_{-k_F}^{k_F}  \!\!\!\!\! dk_2 
\frac{1}{ q ( q + k_1-k_2) r (r + k_1-k_2) }  \nonumber\\
-2
\int_{-\infty}^\infty dq 
\int_{-\infty}^\infty dr 
\int_{k_F}^{\infty} d|k_1|  
\int_{-k_F}^{k_F} dk_2 
\frac{1}{ q r (k_1+k_2)^2 }  
\Big)  ,
\nonumber\\
\label{echd3int}
\nonumber \\  
\eea 
with $|k_1+q|>k_F$, $|k_1+r|>0$,  $|k_2-q|>k_F$, and $|k_2-r|>0$ for the first term,
with $|k_1+q|<k_F$, $|k_1+r|<0$,  $|k_2-q|<k_F$, and $|k_2-r|<0$ for the second term,
and 
with $|k_1+q|<k_F$, $|k_1+r|<0$,  $|k_2-q|>k_F$, and $|k_2-r|>0$ for the final term.  The notation $\int_{k_F}^{\infty} d |k|  $ stands for the sum of two integrals, $\int_{k_F}^{\infty}  dk +  \int_{-\infty}^{-k_F}  dk $.
$k_1$ and  $k_2$ are particle (or hole)  momenta, and $q$ and $r$ are the  momentum transfers.  $n_s$ is the number of spin species, in this case 2.  The limits of integration and constraint inequalities ensure that 
particle  have less momentum than the Fermi-momentum, and
holes have higher momentum than the Fermi momentum.
There is a symmetry factor of $1/2$ associated with each diagram.  
The third diagram contributes twice because a horizontal rotation produces a new diagram with the same numerical value.   This third diagram contributes a negative value because a odd number of vertices connect particle to holes.  To solve Eq. (\ref{echd3}) exactly, we re-scale as follows: 
$q   = k_F v$, 
$r   = k_F x$, 
$k_1 = k_F y$, 
and
$k_2 = k_F z$.  The correlation energy per particle in third order in $\lambda$ is found to be 
\bea
\epsilon_C^{(3)} =  \frac{\lambda^3}{16\pi^4}
\left(\frac{L}{N}\right) \; 
\left(
{\cal I}_a  + {\cal I}_b  - 2 {\cal I}_c \right)
= \frac{\zeta(3)}{2\pi^4} \frac{\lambda^3}{n}\nonumber\\
\label{echd3}
\eea
using the quadrature results, ${\cal I}_a =$
\bea
2 \!
  \int_2^\infty \!\! dv
  \int_2^\infty \!\! dx
  \int_{-1}^1 \! dy
  \int_{-1}^1 \! dz
  \frac{1}{vx (v+y-z) (x+y-z) } \nonumber \\
+ 2\!\! 
  \int_2^\infty \!\! dv
  \int_{-\infty}^{-2} \!\!\! dx
  \int_{-1}^1 \! dy
  \int_{-1}^1 \! dz
  \frac{1}{vx(v+y-z)(x+y-z)} \nonumber \\
+ 4\!\! 
  \int_0^2 \!\!  dv
  \int_0^v \!\!  dx
  \int_{1-x}^1 \!\!\! dy
  \int_{-1}^{-1+x} \!\!\! dz
  \frac{1}{vx (v+y-z) (x+y-z) } \nonumber \\
+ 2\!\! 
  \int_0^2 \!\! dv
  \int_{2-v}^2 \!\! dx
  \int_{1-v}^{x-1} \!\!\! dy
  \int_{1-x}^{v-1} \!\!\! dz
  \frac{1}{vx (v+y-z) (x-y+z) } \nonumber \\
+ 4\!\!
  \int_2^\infty \!\! dv
  \int_0^2 \!\!  dx
  \int_{1-x}^1 \!\!\! dy
  \int_{-1}^{-1+x} \!\!\! dz
  \frac{1}{vx (v+y-z) (x+y-z) } \nonumber \\
+ 4\!\! 
  \int_2^\infty \!\! dv
  \int_{-2}^0 \!\! dx 
  \int_{-1-x}^{-1} \!\!\! dy
  \int_{1+x}^{1} \!\!\! dz
  \frac{1}{vx (v+y-z) (x+y-z) } \nonumber \\
= 8\pi^2 \ln 2 - 36 \; \zeta(3)  ,\nonumber \\
\label{quadrature_a}
\eea 
 ${\cal I}_b =$
\bea
4 \!
  \int_2^\infty \!\! dv
  \int_{v-2}^v  \!\! dx
  \int_{v-1}^{1+x} \!\!\! dy
  \int_{v-1}^{1+x} \!\!\! dz
  \frac{1}{vx (y+z-v)(y+z-x)} \nonumber \\
+ 4 \! \! 
  \int_0^2 \!\! dv
  \int_0^v \!\! dx
  \int_1^{1+x} \!\!\! dy
  \int_1^{1+x} \!\!\! dz
  \frac{1}{vx (y+z-v) (y+z-x) } \nonumber \\
= - 8\pi^2 \ln 2 + 48 \; \zeta(3) , \nonumber \\
\label{quadrature_b}
\eea
and  ${\cal I}_c =$
\bea
4 
  \int_0^2 \! dv
  \int_0^v \! dx
  \int_{1}^{1+x} \!\! dy
  \int_{1-x}^{1} \!\! dz
  \frac{1}{vx(y+z)^2 } \nonumber \\
+4 \!\!
  \int_2^4 \!\! dv
  \int_2^v \!\! dx
  \int_{v-1}^{1+x} \!\!\! dy
  \int_{-1}^{1} \!\! dz
  \frac{1}{vx(y+z)^2 } \nonumber \\
+4 \!\!
  \int_2^4 \! dv
  \int_{v-2}^v \!\! dx
  \int_{v-1}^{1+x} \!\!\! dy
  \int_{1-x}^{1} \!\!\! dz
  \frac{1}{vx(y+z)^2 } \nonumber \\
+ 4 \!\!
  \int_4^\infty \!\! dv
  \int_{v-2}^v \!\! dx
  \int_{v-1}^{1+x} \!\!\! dy
  \int_{-1}^1 \!\!\! dz
  \frac{1}{vx(y+z)^2 } \nonumber \\
= 2\; \zeta(3) . \nonumber \\
\label{quadrature_c}
\eea

The author suspects that an even more tedious 
calculation reveals  that the correlation energy per particle to fourth order in $\lambda$ is 
\bea
\epsilon_C^{(4)} = -\frac{3 \zeta(3)}{4\pi^6} \frac{\lambda^4}{n^2} = -0.000 937 \frac{\lambda^4}{n^2} .
\label{echd4}
\eea
This value agrees well with the numeric Bethe-Ansatz result of -0.00094 $\lambda^4/n^2$.  The full derivation will be given in a later work.


\appendix
  \renewcommand{\theequation}{C-\arabic{equation}}
  \setcounter{equation}{0}  
  \section*{APPENDIX B}  

In the (4,4) Pad{\'e} parameterization of $\epsilon\xc(n,\zeta)$, we have used the result that  there exists an astonishing non-linear analytic  relationship between the given expansion limits and the desired Pad{\'e} parameters.  
A (4,4) Pad{\'e} of the form
\bea
F(x) = \frac{A x^4+B x^3 + C  x^2}{D x^3 + E x^2 + F x +1}
\label{pade}
\eea
has parameters $A$, $B$, $C$, $D$, $E$,and $F$ chosen to satisfy known limits. 
If three terms in both the large and small $x$ limits are known as given below, we can fit these parameters exactly. Specifically, if for large $x$,
\bea
F(x) = g_1 x +g_2+g_3/x +. . . \label{hd}
\eea
and if, for small $x$,
\bea
F(x)=g_4 x^2+g_5 x^3 +g_6 x^4 +. . . \label{ld}.
\eea
the parameters A-F can be determined explicitly.

Let us introduce the following nonlinear Ansatz for the parameters.
\bea
A= \frac{g_1 \left(g_4^3+2 g_1 g_5 g_4+g_2 g_6 g_4-g_2 g_5^2+g_1^2
   g_6\right)}{g_1^3+2 g_2 g_4 g_1+g_3 g_5 g_1+g_3 g_4^2-g_2^2
   g_5},  \nonumber\\ \label{A}
   \eea
\bea
B=    \frac{g_5 g_1^3+g_4^2 g_1^2}{g_1^3+2 g_2 g_4 g_1+g_3 g_5 g_1+g_3 g_4^2-g_2^2
   g_5}  \nonumber\\ 
   + \frac{\left(g_2 g_4 g_5+g_3 \left(g_5^2-g_4 
   g_6\right)\right) \!g_1\! +\! g_2 \! \left(g_4^3+g_2 g_6 g_4-g_2 
   g_5^2\right)}{g_1^3+2 g_2 g_4 g_1+g_3 g_5 g_1+g_3 g_4^2-g_2^2
   g_5} , \nonumber\\ \label{B}
   \eea
\bea
 C=g_4, \label{C}
\eea
\bea
D = A/g_1, \label{D}
\eea
\bea
E=  \frac{g_5 g_1^2+g_4^2 g_1-g_2 g_6 g_1+g_3 g_5^2-g_2 g_4 
   g_5-g_3 g_4 g_6}{g_1^3+2 g_2 g_4 g_1+g_3 g_5 g_1+g_3 
   g_4^2-g_2^2 g_5} , \nonumber\\  \label{E}
 \eea
and
 \bea
 F=\frac{g_4 g_1^2-(g_2 g_5+g_3 g_6) g_1+g_2 g_4^2-g_3 g_4 
   g_5+g_2^2 g_6}{g_1^3+2 g_2 g_4 g_1+g_3 g_5 g_1+g_3 
   g_4^2-g_2^2 g_5} ,\nonumber\\  \label{F}
\eea

Direct substitution of Eqs. \ref{A} to \ref{F} into the Eq. \ref{pade} gives the desired high and low density expansions, Eqs. \ref{hd} and \ref{ld}.


\appendix
  \renewcommand{\theequation}{B-\arabic{equation}}
  \setcounter{equation}{0}  
  \section*{APPENDIX C}  

The exact spin-dependence of the exchange correlation energy was alluded to in the body of the paper.  Here, we present the exact result for the high-density limit.   

For completeness, recall that the spin-dependent Hartree and exchange energy per particle is 
\begin{eqnarray*}
\epsilon_{HX}(n_{\uparrow} ,n_{\downarrow})  
= \lambda n_{\uparrow} n_{\downarrow}  / (n_{\uparrow} +n_{\downarrow}).
\label{ehxhd1}
\end{eqnarray*}

In the high-density limit, correlation energy contributes to second order in $\lambda$.  This contribution is described by the
two-bubble  diagram shown  second  in Figure 7 of Ref. \cite{MB4}.  From  the
standard rules  of perturbation theory, the diagram can be expressed as an integral,
\begin{eqnarray*}
N  \epsilon_C^{(2)}(n_{\uparrow} ,n_{\downarrow}) = \nonumber \\
-\frac{\lambda^2}{L^2}  \frac{L^3}{8\pi^3}  \int_{-\infty}^\infty
dq\; \int_{-k_{F\downarrow}}^{k_{F\downarrow}} dk_1\;  \int_{-k_{F\uparrow}}^{k_{F\uparrow}} dk_2\; \frac{1}{q ( q
+ k_1-k_2)}  , \label{echd1spin}
\nonumber \\
\end{eqnarray*}
with 
$k_{F\uparrow}=n\up/2$,
$k_{F\downarrow}=n\dn/2$,
$|k_1+q|>k_{F\uparrow}$  and $|k_2-q|>k_{F\downarrow}$,
$k_1$ and  $k_2$ are particle  momenta, and $q$ is the  momentum transfer.  Once again the symmetry factor of 1/2 is canceled by the two possible spin configurations.  To solve Eq. (\ref{echd1spin}) exactly, we define two quantities $k_F$ and $s$ according to the following:
$k_{F\uparrow}=k_F (1-s)$ and $k_{F\downarrow}=k_F s$.
Then, we re-scale the coordinates as follows:
$q   = k_F x$,
$k_1 = k_F y$,
and
$k_2 = k_F z$.   
Note that $s=n\dn/n$ and $\zeta=1-2s$.  
After some algebra, we find the correlation energy per particle:
\begin{eqnarray*}
\epsilon_C^{(2)} =  -\frac{\lambda^2}{8\pi^3}\left(\frac{L}{N}\right)
\frac{\pi}{2}\; n\;{\cal I}(\zeta) 
 = -\frac{\lambda^2}{24} f(\zeta) 
\\\nonumber
 = -\frac{\lambda^2}{4\pi^2} \nonumber \\
 \Big(\frac{\pi^2}{2}\!-\!(1\!-\!\zeta)\dilog(1\!-\!\zeta)-(1\!+\!\zeta)\dilog(1\!+\!\zeta) \Big)
\label{echd2}
\end{eqnarray*}
using the quadrature result below and replacing $s$ by $1/2(1-\zeta)$,
\begin{eqnarray*}
{\cal I}(s) =
 \int_{2-2s}^\infty dx\;
  \int_{-s}^s dy\;
  \int_{s-1}^{1-s} dz\;
  \frac{1}{x (x+y-z)} \nonumber\\
 +
   \int_0^{2s} dx\;
  \int_{s-x}^s dy\;
  \int_{s-1}^{x+s-1} dz\;
  \frac{1}{x (x+y-z)} \nonumber\\
+
 \int_{2s}^{2-2s} dx\;
  \int_{-s}^s dy\;
  \int_{s-1}^{x+s-1} dz\;
  \frac{1}{x (x+y-z)} \nonumber\\
= 
4 \left(\frac{\pi^2}{2}-(2-2s)\dilog(2-2s)-2s\dilog(2s) \right)
.\nonumber\\
\label{quadrature}
\end{eqnarray*}


The approximation, $f(\zeta)\approx (1-\zeta^2 )$, is only  true for the extreme values of $\zeta$ with relative errors of up to about 33.333\%.  An expansion of $\epsilon_C$ about $\zeta=0$ for example reveals logarithmic dependencies.  
In the small $\zeta$, nearly unpolarized limit,
\bea
f(\zeta) = 1-\frac{9}{\pi^2}\zeta^2+\frac{6}{\pi^2}\zeta^2 \log \zeta + {\cal O} (\zeta^3)  .
\eea
Perhaps this interesting behavior has implications for the effects of correlation on 1D spin density waves, and this will be explored further in latter work.   Likewise, expansion about $\zeta=1$ shows
\bea
f(\zeta)= 
\frac{6}{\pi^2} \dilog (2\zeta-2) 
+{\cal O} ((\zeta-1)^3)
.
\eea



\appendix
  \renewcommand{\theequation}{D-\arabic{equation}}
  \setcounter{equation}{0}  
  \section*{APPENDIX D}  

The problem in Eq. \ref{1dh2} can be solved analytically within the restricted exact-exchange density functional approach.   For the ground-state, the solution is spin-unpolarized.
The relevant KS equation for one spin-wave-function is
\bea
-\half \nabla^2 \phi(x)
+ \lambda  |\phi(x)|^2 \phi(x)
- Z \sum_{i=1,\pm}^2 \delta(x_i\pm a) \phi(x)
= -\epsilon\phi(x) .
\label{ksequation}
\eea
This is the non-linear Schr{\"o}dinger equation in a double-well potential.  
For simplicity, we only present results for $a=1$.  The solution, vanishing at a distance, can be shown analytically to be
\begin{eqnarray}
\phi(x) = \left\{ \begin{array}{lr}
\sqrt{\frac{1-m}{2m-1}} M \mbox{JacobiNC} (\sqrt{\frac{2\epsilon}{2m-1}} \; x\;,\;m) & |x|<1 \\
M \csch(\sqrt{2\epsilon}(|x|-1) + x_0) & |x|>1  \\
          \end{array} \right.
\end{eqnarray}
where $ \mbox{JacobiNC}$ is a Jacobi Elliptic function.  The double-well potential forces cusps at the $\pm 1$.  
The constraint can be expressed as a transcendental equation.  The other constraint is that the wave-function must normalized to unity.  These two are solved numerically simultaneously.
We present some representative numeric solutions in Table \ref{t:exx}.  These values and others were used in the paper to validate the LDA code.

\begin{table} 
\begin{center}
\begin{tabular}{lllll}\hline
$\lambda$  & $\epsilon$ & $m$   & $M$         & $x_0$     \\ \hline
0.0        & 0.61478    & 1.0   & $\infty$    &  $\infty$ \\ 
0.00068732 & 0.61459    & 0.999 & 42.2794     & 4.77927   \\
0.626582   & 0.44646    & 0.9   & 1.19376     & 1.31123   \\ 
1.182951   & 0.31075    & 0.8   & 0.72483     & 0.93000   \\
1.720499   & 0.19480    & 0.7   & 0.47586     & 0.67034   \\
2.294671   & 0.09239    & 0.6   & 0.28378     & 0.12800   \\
2.63967    & 0.04511    & 0.55  & 0.18487     & 0.29433   \\ 
3.04007    & 0.00886    & 0.51  & 0.07634     & 0.12800   \\
\hline
\end{tabular}
\caption{\label{t:exx} Sample exact-exchange DFT results through the solution of Eq. \ref{ksequation} for a=1 and Z=1.}
\end{center}
\end{table}




\begin{thebibliography}{0} 

\addcontentsline{toc}{section}{References}

\bibitem{CML6}
J. K. Chin, D. E. Miller, Y. Liu, C. Stan, W. Setiawan, C. Sanner, K. Xu and W. Ketterle,  Nature, {\bf 443}, 961 (2006).

\bibitem{SMG6}
T. Stoeferle, H. Moritz, K. Guenter, M. Kohl, and T. Esslinger, \PRL {\bf 96}, 030401 (2006).

\bibitem{MSG5}
H. Moritz, T. Stoeferle, K. Guenter, M. Kohl, and T. Esslinger, \PRL {\bf 94}, 210401 (2005).


\bibitem{MB4}
R.J. Magyar and Kieron Burke, \PRA {\bf 70}, 032508 (2004).

\bibitem{XPAT6}
G. Xianlong, M. Polini, R. Asgari, and M.P. Tosi,
\PRA {\bf 73},  033609 (2006).

\bibitem{KZ4}
Y.E. Kim and A.L. Zubarev, \PRA {\bf 70}, 033612 (2004).



\bibitem{ABGP4}
G. E. Astrakharchik, D. Blume, S. Giorgini, and L. P. Pitaevskii,
\PRL {\bf 93}, 050402 (2004).

\bibitem{BBD6}
R. Benguria, R. Bummelhuis, P. Duclos, S. Perez-Oyarzun, and P. Vytras,
Few-Body Systems {\bf 38}, 133 (2006).

\bibitem{F4}
M.M. Fogler,  \PRL {\bf 94}, 056405 (2005).


\bibitem{NB6}
N.A. Nguyen and A.D. Bandruk, \PRA {\bf 73}, 032708 (2006).

\bibitem{G97}
A. Gold, \PRB {\bf 55}, 9470 (1997).


\bibitem{solvedft}
R.J. Magyar and K. Burke, distributed notes (2000).


\bibitem{R71}
C. Rosenthal, \JCP {\bf 55}, 2474
(1971).


\bibitem{CDR6}
H.D. Cornean, P. Duclos, and B. Ricaud,
Few-Body Systems {\bf 38}, 125 (2006).

\bibitem{HK64} 
P.Hohenberg and W. Kohn, 
\PR {\bf 136}, B864 (1964).

\bibitem{KS65}
W. Kohn and L. J. Sham, \PR {\bf 140}, A1133 (1965).

\bibitem{G67}
M. Gaudin, Phys. Lett. {\bf 24A}, 55 (1967). 

\bibitem{Y67}
C. N. Yang, Phys. Rev. Lett. {\bf 19}, 1312 (1967).

\bibitem{FB80}
W. I. Friesen and B. Bergersen, J. Phys. C {\bf 13}, 6627 (1980).

\bibitem{RFZ3}
A. Recati, P.O. Fedichev, W. Zwerger, and RP. Zoller, J. Opt. B Quantum Semiclass. Opt. {\bf 5}, S55 (2003).

\bibitem{RG84}
E. Runge and E.K.U. Gross, \PRL {\bf 52}, 997 (1984).

\bibitem{C95}
M.E. Casida,  
in \emph{Recent Advances in Density Functional Methods, Part I}, Ed. by Ed. Chong (Singapore, World Scientific, 1995), p. 155.

\bibitem{PSB95}
J.P. Perdew, A. Savin, and K. Burke, \PRA {\bf 51}, 4531 (1995).

\bibitem{AAL97}
V. I. Anisimov, F Aryasetiawan, and A I Lichtenstein, 
J. Phys.: Condens. Matter {\bf 9}, 767 (1997). 

\bibitem{SK0}
S.Y. Savrasov and G. Kotliar, \PRL {\bf 84}, 3670 (2000).

\bibitem{ZGJB97}
P. Ziesche, O. Gunnarsson, W. John, and H. Beck, \PRB {\bf 55}, 10270 (1997).

\bibitem{G57}
J. Goldstone, 
Journal Proceedings of the Royal Society of London. Series A, Mathematical and Physical Sciences {\bf 239}, 1217 
(1957).



\end{thebibliography}
\end{document}